\documentclass[twocolumn,12pt]{article}

\usepackage{graphicx}
\usepackage{dcolumn}
\usepackage{bm}
\usepackage{multirow,multicol}
\usepackage{amsmath,amssymb}
\def\avg#1{\langle #1\rangle }
\def\bx{{\bf x}}
\def\bq{{\bf q}}
\def\bR{{\bf R}}
\def\bD{{\bf D}}
\def\lp{\left(}
\def\rp{\right)}

\usepackage[sort,square,numbers,compress]{natbib}
\setcitestyle{square}

\onecolumn

\begin{document}

\title{\vspace{-1in}\bf\Large First-principles prediction of high-entropy-alloy stability}


\author{R. Feng and P. K. Liaw\thanks{Department of Materials Science and Engineering, University of Tennessee, Knoxville, TN  37996, USA},
M. C. Gao\thanks{National Energy Technology Laboratory, Albany, OR 97321, USA; AECOM, P.O. Box 1959, Albany, OR 97321, USA}, and
M. Widom\thanks{Department of Physics, Carnegie Mellon University, Pittsburgh, PA  15213, USA. Correspondence and requests for materials should be addressed to M.W. (email: widom@cmu.edu).}}

\maketitle

\renewcommand{\baselinestretch}{2}

\begin{abstract}
High entropy alloys (HEAs) are multicomponent compounds whose high configurational entropy allows them to solidify into a single phase, with a simple crystal lattice structure.  Some HEA's exhibit desirable properties, such as high specific strength, ductility, and corrosion resistance,  while challenging the scientist to make confident predictions in the face of multiple competing phases.  We demonstrate phase stability in the multicomponent alloy system of Cr-Mo-Nb-V, for which some of its binary subsystems are subject to phase separation and complex intermetallic-phase formation.  Our first-principles calculation of free energy predicts that the configurational entropy stabilizes a single body-centered cubic (BCC) phase from T = 1,700K up to melting, while precipitation of a complex intermetallic is favored at lower temperatures.  We form the compound experimentally and confirm that it forms as a single BCC phase from the melt, but that it transforms reversibly at lower temperatures.
\end{abstract}

\clearpage
\section{Introduction}
Exploring new materials with outstanding properties is an eternal pursuit of scientists and engineers.  High entropy alloys (HEAs) constitute a newly-emerging class of materials that offer enhanced mechanical and corrosion properties for practical applications, while presenting challenges to the scientist making predictions in the face of compositional complexity~\citep*{Yeh04_1,Cantor04,Zhang08,Zhang2014,Gludovatz14,Li16,Miracle2017,Senkov15,Gludovatz16,Lim16,Lu16,Troparevsky15,Santodonato15}.  Alloys of two or three species often phase-separate, and when they form a compound, it is often a complex ordered structure.  Thus, HEAs are remarkable for forming single phases that are, moreover, simple lattice structures. Trial-and-error experimental approaches to test the phase stability of HEAs are costly and time-consuming, while previously-reported phase formation rules are empirical and susceptible to the alloy systems~\citep{Feng16,Gao17}, motivating a first-principles theoretical approach to accelerate the design of HEAs and control their properties through the prediction of phase stability.

Here we illustrate these ideas by demonstrating phase stability in the quaternary Cr-Mo-Nb-V alloy system. Some of its binary and ternary subsystems are subject to phase separation and complex intermetallic phase formation.  Utilizing fully first-principles calculations of the free energies of the full quaternary and its competing binary and ternary phases, we predict that the configurational entropy stabilizes a single body-centered cubic (BCC) solid solution at high temperatures, but that it will precipitate a complex intermetallic Laves phase at lower temperatures.  Experimentally, we verify formation of the HEA from the melt, confirm that precipitation of the Laves phase at low temperatures, and observe that it reverts to a single BCC phase upon annealing at high temperatures, thus illustrating the entropic stabilization principle of HEAs.

\section{Results}
\subsection{Quaternary alloy system}
\begin{figure*}
\includegraphics[clip,width=6.0in]{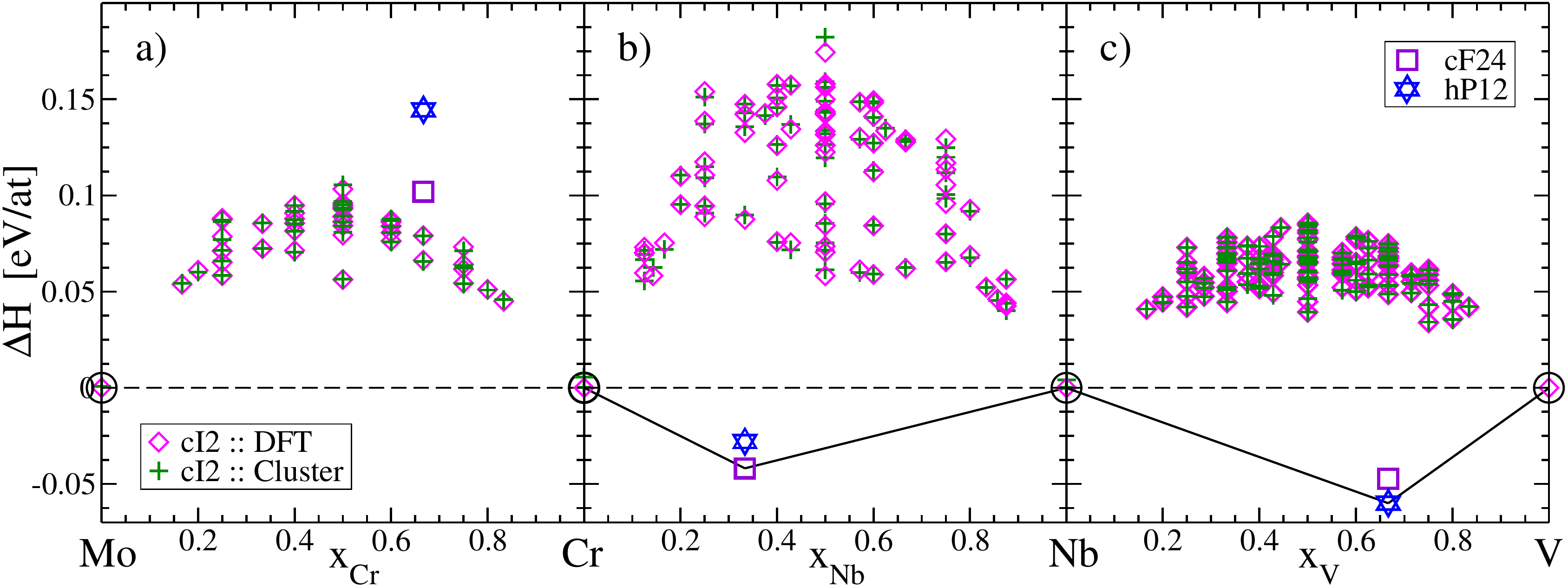}
\caption{\label{fig:3binaries}Formation enthalpies calculated from DFT (magenta diamonds) and cluster expansions~\citep{Walle02a,WidomHEABook} (green plus signs) for (a) Cr-Mo, (b) Cr-Nb, and (c) Nb-V. Violet square and blue star are Laves Phases of Pearson types cF24 (Strukturbericht C15) and hP12 (C14), respectively. Solid black lines are convex hulls.}
\end{figure*}

Consider the four-component alloy system of Cr-Mo-Nb-V.  All four elements take body-centered-cubic (BCC, Pearson type cI2, Strukturbericht A2) structures.  Taking elements pairwise, we note that mixing enthalpies for the BCC solid solution are positive for Cr-Mo, Cr-Nb, and Nb-V (see Fig.~\ref{fig:3binaries}), indicating low-temperature phase separation, while the remaining cases, Cr-V, Mo-Nb, and Mo-V, have negative enthalpies~\citep{Gao08,Jiang04}, indicating stability down to low temperatures.  Experimentally, Cr-Nb has a complete miscibility gap at all temperatures below melting, while phase separation has not been seen experimentally in the case of Cr-Mo~\citep{Kocherzhinskii85} and Nb-V~\citep{Smith90}, though it {\em is} predicted~\cite{Turchi06,Molokanov77}.  Cr-Nb and Nb-V are predicted~\citep{WidomHEABook} to form Laves phases of types cF24 (Strukturbericht C15) and hP12 (Strukturbericht C14).  Both Laves phases are observed experimentally in Cr-Nb~\citep{Blazina86}, while neither has yet been reported in Nb-V.  Experimental investigations of the ternaries show that Cr-Mo-Nb phase-separates~\citep{Svechnikov64} and Cr-Nb-V forms a C15-Laves phase~\citep{Takasugi95}, while Cr-Mo-V and Mo-Nb-V become complete solid solutions~\citep{Kocherzhinskii85}.  No quaternary phases have been reported experimentally.

Inspection of Fig.~\ref{fig:3binaries} offers a clue to why experiments on Cr-Mo and Nb-V fail to observe the predicted miscibility gaps, while separation {\em is} seen in Cr-Nb.  The mixing enthalpies are relatively low in Cr-Mo and Nb-V, compared with Cr-Nb, so their estimated critical temperatures for phase separation are low (1,213K and 653K, respectively~\citep{Turchi06,Molokanov77}).  Meanwhile, the melting temperatures of Cr-Mo and Nb-V are relatively high (2,093K and 2,133K, respectively, compared with 1,893K for Cr-Nb).  Since atomic diffusion far below the melting temperature can be slow, phase separation will be difficult to observe.  The problem is even more severe for complex structures, potentially explaining the failure so far to observe the predicted NbV$_2$ C14-Laves phase.

Turning to the full quaternary, we will test its stability at high temperatures by calculating its free energy relative to the most-likely competing phases, namely, the coexistence of a Mo-rich BCC medium entropy alloy with a CrNbV-rich Laves phase.  A working hypothesis is that the single phase is stabilized by a large configurational entropy of mixing, and that the entropy is maximized for the simple BCC lattice structure in which all sites are equivalent.  However, we expect the lower enthalpy of the competing phases to predominate at lower temperatures where the entropy is less relevant, leading to precipitation of a ternary CrNbV Laves phase as temperature drops.  At very low temperatures, we anticipate complete phase separation into four coexisting phases.

\subsection{Free energy calculation}
For each phase, we calculate the Helmholtz free energy as a sum of discrete configurational, vibrational, and electronic contributions~\citep{Fultz10,WidomHEABook}, $F(\bx,T,V)=F^c+F^v+F^e$. The composition variable, $\bx$, is expressed as a four-component vector in our composition space.  Because metallurgical experiments are generally carried out at low pressure ($P\approx 0$), the Gibbs free energy, $G(\bx,T)=\min_V(F+PV)$, of each phase is simply the Helmholtz free energy, $F$, evaluated at its minimizing volume.  If we have a mixture of phases, $i$, each having a mole fraction $\gamma_i$, the total free energy is $G=\sum_i \gamma_i G_i(\bx_i,T)$.  

Consider the phase separation of a CrNbV Laves phase from a CrMoNbV BCC solid solution, expressed as
\begin{equation}
\label{eq:react}
{\rm Cr}{\rm Mo}{\rm Nb}{\rm V}\rightarrow
{\rm Cr}_{1-\gamma}{\rm Mo}_{1}{\rm Nb}_{1-\gamma}{\rm V}_{1-\gamma} + 
{\rm Cr}_\gamma{\rm Nb}_\gamma{\rm V}_\gamma,
\end{equation}
where $\gamma\in(0,1)$ represents the progress of the reaction.  The fraction of the Laves phase is $3\gamma/4$, leaving a fraction $1-3\gamma/4$ of the BCC phase. The BCC phase becomes increasingly Mo-rich, with the composition of $x_{\rm Mo}=1/(4-3\gamma)$, while the remaining elements have the composition $x_\alpha=(1-\gamma)/(4-3\gamma)$ for $\alpha = $ Cr, Nb, or V.  Thus, the total free energy for a two-phase mixture of Laves with a partially-transformed BCC structure is
\begin{align}
\label{eq:Gtot}
G_{\rm Tot}(x_{\rm Mo})
   &= (1-3\gamma/4)G_{\rm BCC} + (3\gamma/4)G_{\rm Laves}\\ \nonumber
   &= G_{\rm Laves}+(G_{\rm BCC}-G_{\rm Laves})/4x_{\rm Mo},
\end{align}
and $G_{\rm BCC}$ depends on $x_{\rm Mo}$.  Equation~(\ref{eq:Gtot}) holds for the total Gibbs free energy and also for the separate configurational, vibrational and electronic components.

To calculate the configurational free energy, $G_c(\bx,T)$, we apply our hybrid Monte Carlo/molecular dynamics method (MCMD~\citep{Widom13}) to anneal the short-range chemical order.  Over the temperature range of interest (T = 1,000K through melting around 2,133K), the chemical order was found to be moderate and nearly temperature independent.  In particular, the configurational entropy, remained within 2\% of the ideal value~\citep{Widom16}.  Thus we take the mixing entropy of the BCC solid solution as ideal, $S^c=k_B \sum_\alpha x_\alpha \log{x_\alpha}$.  We calculate the configurational enthalpy of formation, relative to pure elements, $\Delta H_c$, by quenching our simulated structures to T = 0K through the relaxation of atomic coordinates and lattice parameters.  The resulting enthalpies (see Fig.~\ref{fig:x-series}a) are nearly independent of the annealing temperature but exhibit a strong variation with respect to composition.

Applying MCMD to the C15-Laves phase of CrNbV reveals that Cr and V readily substitute on the $16d$ Wyckoff sites but they do not mix to any appreciable degree with Nb, which remains on the $8a$ sites.  Hence, we take the entropy of CrNbV as $S^c=(2/3) k_B \log{2}$.  The total Gibbs configurational free energy of formation is the sum of the BCC and Laves terms weighted by their respective fractions, as in Eq.~(\ref{eq:Gtot}).

\begin{figure}
\includegraphics[clip,width=6.0in]{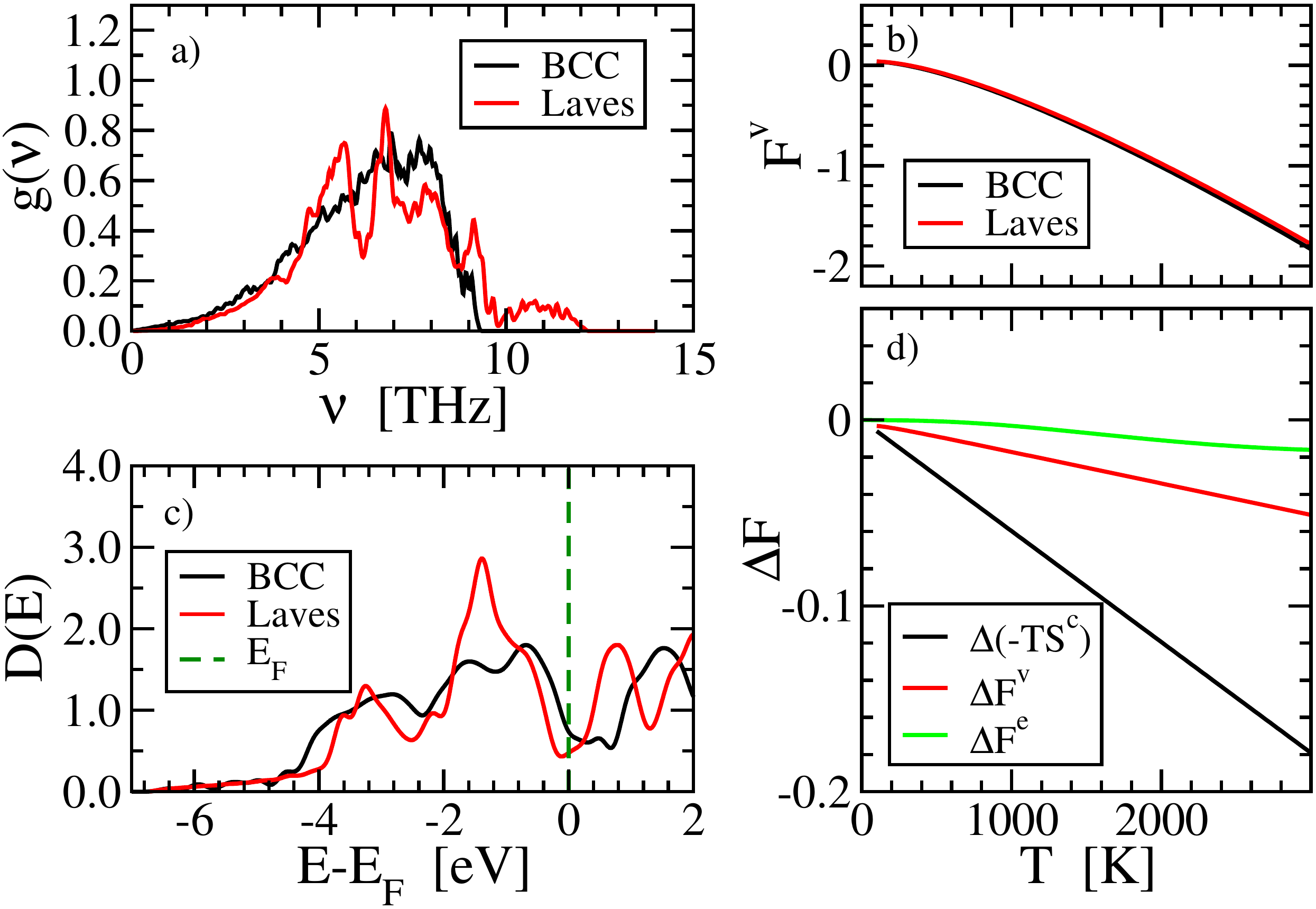}
\caption{\label{fig:4part} Contributions to free energies. The BCC composition is CrMoNbV, while the Laves phase is CrNbV. (a) Vibrational density of states (modes/THz/atom); (b) Vibrational free energies (eV/atom); (c) Electronic density of states (states/eV/atom); (d) Configurational, vibrational, and electronic free energy differences, $F_{\rm BCC}-F_{\rm Laves}$, all favor the BCC phase.}
\end{figure}

The vibrational free energy is calculated in the harmonic approximation from the vibrational density of states, $g(\nu)$, {\em via} the single vibrational mode free energy
\begin{equation}
\label{eq:Fvib}
f^v(\nu) =k_BT \log{[2\sinh{(h\nu/2k_BT)}]},~~~~
F^v =\int g(\nu) f^v(\nu) {\rm d}\nu,
\end{equation}
where $g(\nu)$ is obtained by the direct method~\citep{Parlinski97} by means of force constants that we determine using density functional perturbation theory at T = 0K.  Figure~\ref{fig:4part}a compares vibrational densities of states for the equiatomic BCC phase and the Laves phase.  The Laves phase has some high-frequency phonons that are absent in the BCC phase, while the BCC phase has a small but persistent excess of low-frequency phonons, lowering $F^v_{\rm BCC}$ relative to $F^v_{\rm Laves}$ (Fig.~\ref{fig:4part}c and d).  These differences reflect the tetrahedral close-packed structure of the Laves phase, compared with the relatively-open BCC structure.  To model phase separation, we require the composition dependence of $\Delta F^v$.  We find that the differences in the BCC vibrational free energy with respect to composition, $x_{\rm Mo}$, are small, compared with the difference between the BCC and Laves phases, and hence may be treated as composition-independent.

The electronic free energy is calculated from the single electron density of states, $D(E)$, predicted at T = 0K by density functional theory (DFT), and the Fermi occupation function, $f_T(E)=1/(\exp{((E-E_F)/k_BT)}$, and associated entropy, $s_T(E)$, for individual states of energy $E$,
\begin{equation}
s_T(E) =f_T(E)\log{f_T(E)}+(1-f_T(E))\log{(1-f_T(E))}.
\end{equation}
We determine the electronic free energy, $F^e=E^e-TS^e$, where
\begin{equation}
\label{eq:Fe}
E^e = \int {\rm d}E~D(E) (f_T(E)-f_0(E)) (E-E_F),~~~~
S^e = -k_B\int {\rm d}E D(E) s_T(E).
\end{equation}
Figure~\ref{fig:4part}c shows that the electronic densities of states near $E_F$ differ significantly between BCC and Laves phases.  This difference has a significant composition dependence, as can be seen within a rigid band model from the irregular shape of $D(E)$ near $E_F$ (increasing $x_{\rm Mo}$ increases $E_F$).

\begin{figure}
\includegraphics[clip,width=6.0in]{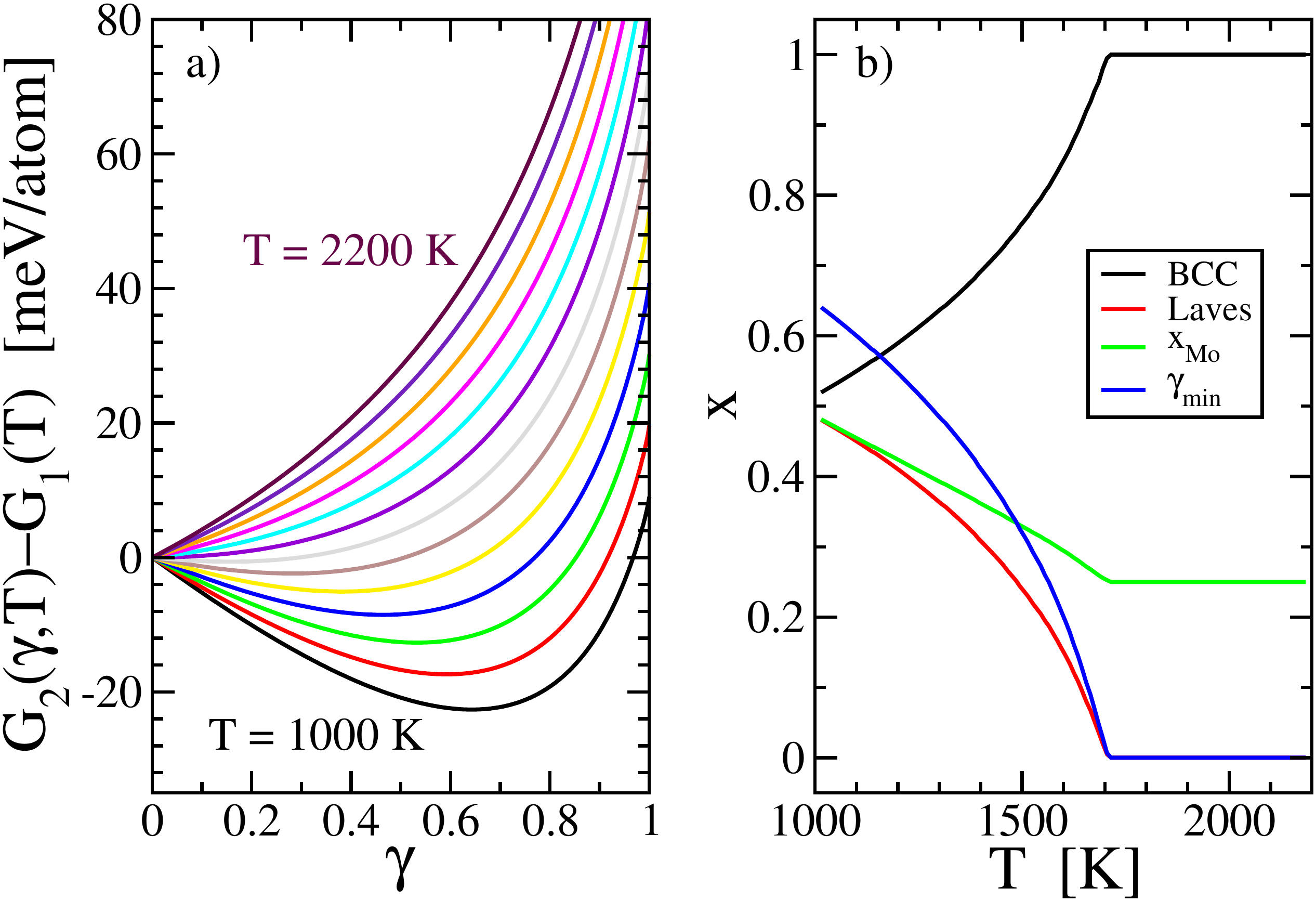}
\caption{\label{fig:Gofx} (a) Free energy, $G_2(\gamma,T)$, for a two-phase BCC + Laves mixture with compositions given by Eq.~(\ref{eq:react}), relative to $G_1$, the free energy of the equiatomic BCC HEA. (b) Minimizing values of $x_{\rm Mo}$, $\gamma$, and BCC and Laves phase fractions as functions of temperature.}
\end{figure}

Because we utilize the harmonic approximation, thermal expansion is neglected, and the vibrational Helmholtz free energies, $F^v$ and $F^e$, can be added directly to the configurational Gibbs free energy, $G^c$.  Thus, combining the above results, we have obtained the free energy, $G$, expressed as a function of transformation, $\gamma$, or composition, $x_{\rm Mo}$, and temperature, T.  Figure~\ref{fig:Gofx} graphs this function, revealing minima with $\gamma>0$ for T$<$1,700K, implying phase separation.  From T = 1,700K up to melting above T = 2,100K, however, the minima occur at the boundary, $\gamma=0$ ({\em i.e.}, $x_{\rm Mo}=1/4$), indicating that a single BCC phase is stable.

\subsection{Experimental results}
\label{sec:exptl}

\begin{figure}
\includegraphics[clip,width=6.0in]{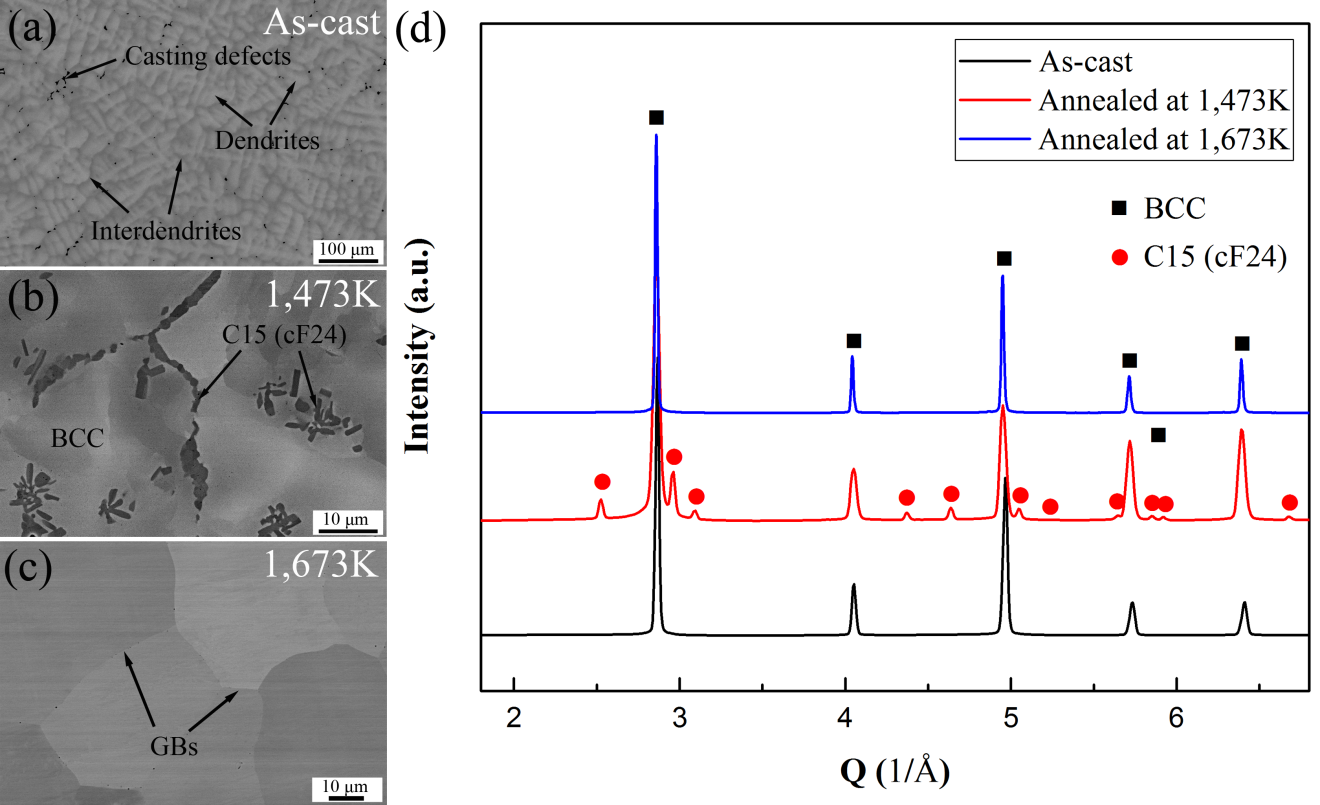}
\caption{\label{fig:expt}  Electron micrographs of the (a) as-cast Cr-Mo-Nb-V sample, (b) following the T = 1,473K anneal, and (c) following the T = 1,673K anneal. (d) Synchrotron x-ray patterns.}
\end{figure}

To test this prediction, we formed a sample by arc melting and casting. The as-cast sample was found to be a single BCC phase, with the characteristic of dendrites and interdendrites (Figs.~\ref{fig:expt} a and d). The single phase remained intact under 21 days annealing at temperatures up to 1,273K, but exhibited precipitation of a C15-Laves phase under annealing of 3 days at 1,473K (Fig. 4b and d). Subsequent low-temperature annealing failed to restore the single phase, indicating that the phase-separated state is indeed stable at low temperature.  The C15 phase disappeared when the 1,473K-annealed sample was annealed at 1,673K for 12 hours (Fig. 4c and d). Meanwhile, equiaxed grains appeared instead of the original dendrites and interdendrites, indicating the pristine BCC phase was restored at 1,673K. This reversible transformation (BCC + Laves phases at low temperatures, and full BCC phase at high temperatures) is consistent with our theoretical prediction.  Further annealing at 1,873K revealed additional impurity-stabilized phases (see Fig.~\ref{fig:sem}b), including a novel CCr$_3$Nb$_3$ carbide of Pearson type cF112.

Table~\ref{tab:edx} reports the compositions of the C15 and BCC phases in the 1,473K-annealed state. Figure~\ref{fig:edx-BCC} shows the distribution of constitutive elements after annealing the 1,473K-annealed sample at 1,673K for 12 hours. No obvious segregation was observed within the BCC phase in the 1,673K-annealed state.  Figure~\ref{fig:sem} presents the microstructures after annealing the alloy at 1,273K and 1,873K. It can be seen that the alloy still keeps the characteristic of dendrites and interdendrites under the 21 days annealing at 1,273K (Fig.~\ref{fig:sem}a). After annealing at 1,873K, multiple phases appeared in the alloy, including BCC, CCr$_3$Nb$_3$, and an unidentified needle-like phase, (Fig.~\ref{fig:sem}b).

\begin{figure}
\includegraphics[clip,width=6.0in]{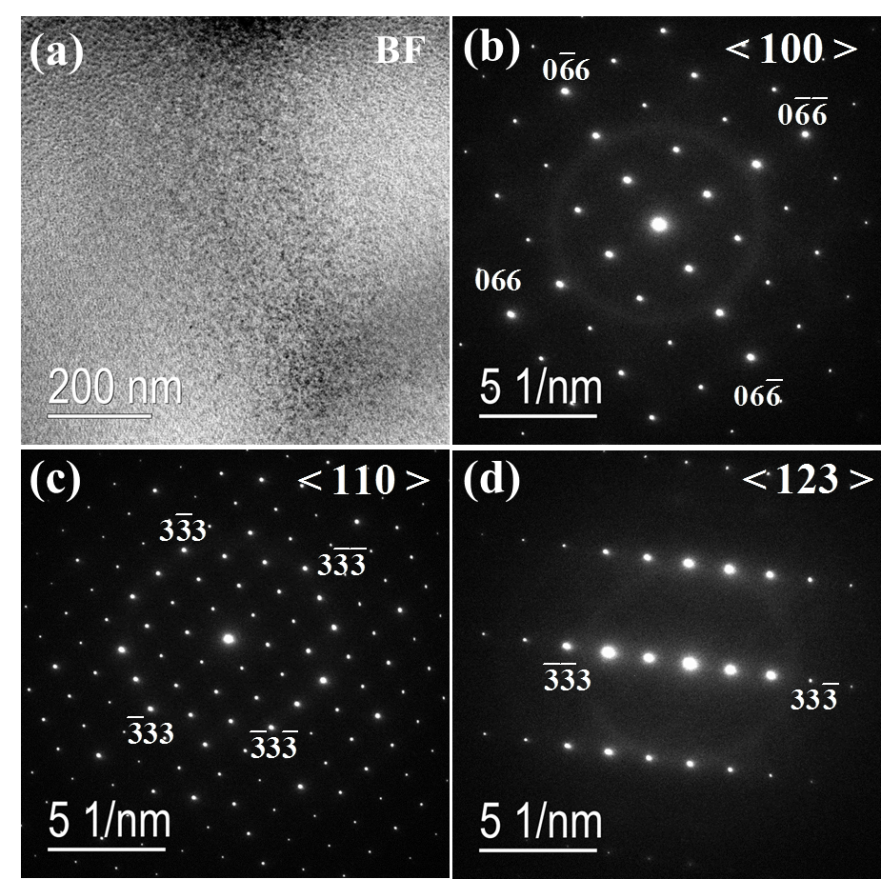}
\caption{\label{fig:tem-carbide} (a) Bright-field (BF) image of the CCr$_3$Nb$_3$ phase in the T = 1,873K-annealed state. (b)-(d) Selected area electron diffraction (SAED) patterns at zone axis $\avg{100}$, $\avg{110}$, and $\avg{123}$.}
\end{figure}

The chemical distribution of the CCr$_3$Nb$_3$ phase in the 1,873K-annealed state was further detected by EDX, as shown in Fig.~\ref{fig:edx-carbide}.  An Fd$\bar{3}$m symmetry was identified by TEM SAED (Fig.~\ref{fig:tem-carbide}), suggesting a phase of the structure type of NiTi$_2$ with a lattice parameter $a=11.7$~\AA. As demonstrated below, this phase is stabilized by carbon impurities and can incorporate silicon substitution in place of Cr.

\section{Discussion}

So far we focused on the primary competitor to the HEA, namely, the separation into a Laves phase plus a BCC medium entropy alloy (MEA).  In principle, according to the Gibbs phase rule, the system could contain up to four independent phases at low temperature.  We employed the Alloy Theoretic Automatic Toolkit cluster expansion~\cite{Walle02a} to generate a variety of candidate low-temperature binary, ternary, and quaternary configurations based on decorations of the BCC lattice.  Based on enthalpies, we predict that the equiatomic CrMoNbV decomposes at low temperatures~\cite{WidomHEABook} into four competing phases, Cr$_2$V.tI6, CrNbV.cF24, MoNb.oC12, and Mo$_4$V$_3$.hR7 subject to the stoichiometric relationship
\begin{equation}
\label{eq:sep}
10~{\rm CrMoNbV}~\rightarrow~
3~{\rm Cr}_2{\rm V}~+~6~{\rm Mo}{\rm Nb}~+~4~{\rm CrNbV}~+~1~{\rm Mo}_4{\rm V}_3.
\end{equation}
Relaxed lattice parameters and Wyckoff coordinates of these structures are given in Table~\ref{tab:Wyckoff}.

Assuming that all four phases are stoichiometric, we neglect the configurational entropy and set $G^c=\Delta H^c$. Combining $G^c$ with $F^v$ and $F^e$, and weighting the contributions according to Eq.~(\ref{eq:sep}), we obtain the function displayed in Fig.~\ref{fig:G-all} labeled as $G_4$.  Notice that $G_4$ crosses the single-phase equiatomic HEA free energy, $G_1$, around T = 800K, indicating a transition from the HEA to the 4-phase mixture at low temperatures.  On the same figure, we plot the free energy, $G_2$, of the 2-phase mixture, and this value lies {\em below} $G_1$ and $G_4$ over the range of T = 500-1,700K.  Accordingly, the HEA does not transform directly to the 4-phase mixture upon cooling, but instead exhibits a 2-phase region in between.  The upper transition at T = 1,700K is continuous, with $G_2$ joining $G_1$ tangentially.  This trend is reflected in the continuity of composition variables, as illustrated in Fig. 3b of the main text.  The lower transition is discontinuous within the present model, with $G_4$ crossing $G_2$ transversally at 500K.  However, because it corresponds to chemical ordering on the BCC lattice, the actual behavior may be more complicated with multiple intervening phases and possibly continuous transitions.  These transitions occur at low temperatures, making it doubtful that they can be observed experimentally.

Because CrNb$_2$ exhibits a high-temperature C14-Laves phase (Pearson type hP12), and the structure is predicted to be stable at low temperatures in NbV$_2$, we checked the free energy of CrNbV as a C14-Laves phase.  At this equiatomic composition, the free energy of C15 remains below the free energy of C14, for all temperatures below melting.

As mentioned above, TEM identified a Cr-Nb-rich impurity phase of space group $Fd\bar{3}m$ (group \#227), and proposed the structure type of NiTi$_2$ (Pearson type cF96).  This structure proved energetically unfavorable as a Cr-Nb binary.  However, carbon impurities at octahedral interstitial sites stabilize the phase in compositions CCr$_3$Nb$_3$ and CCr$_2$Nb$_4$ (both Pearson type cF112).  Thus, we predict the existence of a previously unknown stable carbide in the C-Cr-Nb alloy system.  Table~\ref{tab:carbide} gives crystallographic details of the CCr$_3$Nb$_3$.cF112 phase, which we predict to be stable against all known binary and ternary competing phases, with the formation enthalpy of $\Delta H=-220$~meV/atom, relative to pure elements.  Substituting a single Si atom at a Cr2 site results in a quaternary that is stable against all known competing phases, with the formation enthalpy of $\Delta H=-275$~/meV/atom.   CCr$_2$Nb$_4$ is similar to CCr$_3$Nb$_3$, but with Nb occupying the 16c site.

In summary, we have predicted the existence of a single-phase refractory HEA.  This phase is stabilized at high temperatures by the configurational entropy of mixing despite the immiscibility of constituent elements and a competing complex intermetallic phase, which leads to phase separation at low temperatures.  A reversible phase transformation restores the entropically stabilized BCC phase at high temperatures, illustrating one of the foundational principles of HEAs.  Our theoretical methods, combining the quantum-mechanical total-energy calculation with statistical mechanics to predict free energies, can be applied to many problems in alloy design.

\newpage

\section{Methods}

\subsection{Enthalpy of formation}
First-principles calculations are performed using projector augmented wave potentials~\cite{Blochl94,Kresse99} in the generalized gradient approximation~\cite{Perdew96} as implemented in the Vienna Ab-initio Simulation Package (VASP~\cite{Kresse93}).  Structures were fully relaxed in lattice parameters and internal coordinates at T = 0K using the energy cutoff of 400 eV and $k$-point densities sufficient to achieve convergence of 1 meV/atom or better. Hybrid Monte Carlo/molecular dynamics simulations (MCMD~\cite{Widom13}) generated BCC solid-solution configurations over a series of Mo-rich compositions with an appropriate short-range order.  These calculations used default energy cutoffs and a single $k$-point in systems of size $N=128$ atoms.  Figure~\ref{fig:x-series}a shows the resulting enthalpies of formation, relative to pure elements, which we find fit well to a quadratic polynomial.

\subsection{Electronic free energy}
We utilize the same series of Mo-rich configurations to evaluate electronic densities of states and electronic free energies.  The integral in Eq.~(\ref{eq:Fe})  can be expanded in power series in temperature~\cite{WidomHEABook}.  Expressing $D(E)=D(E_F)+D'(E_F)(E-E_F)+D''(E_F)(E-E_F)^2+\cdots$, we find
\begin{equation}
F^e=-D(E_F)\pi^2 T^2/6-7D''(E_F)\pi^4T^4/180-\cdots.
\end{equation}
Odd powers drop out, and truncating at $T^4$ is sufficient even at quite high $T$.  Here we fit coefficients, $a$ and $b$, to $F^e(T)=a(T/1000)^2+b(T/1000)^4$ over the range of T = 0-2,000K.  The composition dependence of $F^e$ is irregular, as can be inferred from the double minima in $D(E)$ (see Fig.~\ref{fig:x-series}). Hence, we model $F^e$ by fitting $a$ and $b$ to 6th-order polynomials of $x_{\rm Mo}$, as illustrated in Fig.~\ref{fig:x-series}c.

\subsection{Vibrational free energy}
The vibrational free energy is calculated in the harmonic approximation from the vibrational density of states, $g(\nu)$.  This is obtained from the force constant matrix, ${\bf \Phi}_{ij}=\partial^2 U/\partial\bR_i\partial\bR_j$, which VASP determines using density functional perturbation theory at T = 0K.  Our calculations utilize a plane-wave basis with the energy cutoff of 400 eV, ``accurate'' precision to avoid wrap-around errors, and an additional support grid for augmentation charges.

The dynamical matrix at the wavevector $\bq$ is~\cite{Parlinski97,Fultz10}
\begin{equation}
\bD_{ij}(\bq)=\frac{1}{\sqrt{m_im_j}}~e^{2\pi i \bq\cdot\bR_{ij}}{\bf \Phi}_{ij}.
\end{equation}
We diagonalize $\bD_{ij}(\bq)$ to obtain vibrational frequencies, $\nu(\bq)$, and sample the full Brillouin zone to obtain the density of states, $g(\nu)$.  Representative densities of states, calculated in 27-atom cells for various compositions, are shown in Fig.~\ref{fig:vDOS}.  Notice that $g(\nu)$ is relatively insensitive to the composition, but that the BCC DOS differs significantly from the C15-Laves phase.

Because we utilize the harmonic approximation, thermal expansion is neglected, and the vibrational Helmholtz free energy, $F^v$, can be added directly to the configurational Gibbs free energy, $G^c$.  In the high-temperature range of interest, the free energy varies classically, with quantum effects relevant only for establishing offsets in the energy and entropy.  Thus, we can fit $F^v(T)$ with high precision using only a single parameter, an effective Debye temperature, $\Theta_D$, chosen to match the quantum-based free energy in the classical regime.  We set
\begin{equation}
\label{eq:Debye}
F^v(T) = \frac{9}{8}k_B\Theta_D-k_B T\left[D_3(\Theta_D/T)-\log\lp 1-e^{-\Theta_D/T}\rp\right]
\end{equation}
with $D_3(z)=1-3z/8+z^2/20+\cdots$ the third order Debye function~\cite{Moruzzi88}.  We set $\Theta_D$=393K for the BCC HEAs, and $\Theta_D$=420K for the Laves phase, as appropriate values to reproduce the high temperature free energy.

\subsection{Sample preparation}
The CrMoNbV alloy was fabricated by arc-melting the constituent elements (purity $> 99.9$ weight percent), then drop casting into a water-cooled copper hearth. To achieve a homogeneous distribution of elements in the alloy, the melting and solidification processes were repeated five times.  The annealed samples of the CrMoNbV alloy were encapsulated in quartz tubes, filling with high-purity argon after vacuuming several times. The samples were heat treated at 1,273K for 21 days and 1,473K for 3 days, respectively, and then water quenched to validate the stability of the BCC phase. After obtaining the 1,473K-annealed samples, another annealing treatment at 1,673K for 12 hours was performed using a 1,473K-annealed sample to verify the stability of the C15-Laves phase. Further annealing at 1,863K revealed a novel carbide stabilized by impurities.

\subsection{Microscopy}

A LEO Gemini 1525 field emission scanning electron microscopy (SEM) coupled with energy-dispersive X-ray spectroscopy (EDX) was utilized to characterize the microstructure and composition of this alloy. Transmission-electron microscopy (TEM) was conducted to identify the new phase in the 1,863K-annealed state, using ZEISS LIBRA 200 HT FE MC coupled with EDS. SEM specimens were initially polished with 1200-grit SiC paper and, subsequently, vibration polished using 0.05 $\mu$m SiC liquid for the final surface clarification. Focused ion beam (FIB) milling was used for the preparation of TEM specimens, targeting the specific phase.  The average compositions of each phase were measured at five different locations to ensure compositional accuracy.

\subsection{X-ray diffraction}
Synchrotron X-ray diffraction experiments were carried out at the Argonne National Laboratory, Advanced Photon Source, using the beamline 11-ID-C. The bulk specimens ($\sim$1 mm thick) after the SEM characterization were measured by a beam energy of 111 keV (0.11798~\AA) with a beam size of $0.5\times0.5$ mm.

\section{Addendum}
Theoretical calculations at CMU were supported by Department of Energy grant DE-SC0014506 and by the National Science Foundation through XSEDE grant DMR160149 at the Pittsburgh Supercomputer Center.  MCGao was supported through NETL's Office of Research and Development's Innovative Process Technologies (IPT) Field Work Proposal.  RF and PKL very much appreciate the support of the U.S. Army Research Office project number W911NF-13-1-0438. PKL also acknowledges the support from the National Science Foundation under grant DMR-1611180.

RF conducted the experiments; PKL provided the experimental idea to verify the theory; MCG contributed to the experimental design; MW proposed the alloy system and performed free-energy calculations to test the stability.  All authors discussed the results and commented on the manuscript.

The authors declare that they have no competing financial interests.

\newpage


\bibliographystyle{./naturemag}
\bibliography{refs}

\clearpage

\clearpage

\begin{table}[h]
\caption{\label{tab:edx} Chemical composition (atom \%) of observed phases in the T = 1,473K annealed state.}
\begin{tabular}{c|cccc}
Phase &      Cr      &      Mo      &      Nb      &       V     \\
\hline
BCC   & 23.70 $\pm$ 0.4 & 24.48 $\pm$ 0.6 & 24.84 $\pm$ 0.3 & 26.98 $\pm$ 0.3 \\
Laves & 44.82 $\pm$ 0.9 &  7.59 $\pm$ 0.7 & 30.53 $\pm$ 0.5 & 17.06 $\pm$ 0.5
\end{tabular}
\end{table}

\clearpage

\setcounter{equation}{0}
\setcounter{figure}{0}
\setcounter{table}{0}
\setcounter{page}{1}
\setcounter{section}{0}
\makeatletter
\renewcommand{\thepage}{S\arabic{page}}
\renewcommand{\thesection}{S\arabic{section}}
\renewcommand{\theequation}{S\arabic{equation}}
\renewcommand{\thefigure}{S\arabic{figure}}
\renewcommand{\bibnumfmt}[1]{[S#1]}
\renewcommand{\citenumfont}[1]{S#1}

\clearpage
\begin{center}
{\bf \Large Supplemental material for: first-principles prediction of high entropy alloy stability}
\end{center}

\begin{figure*}[h]
\includegraphics[clip,width=1.8in]{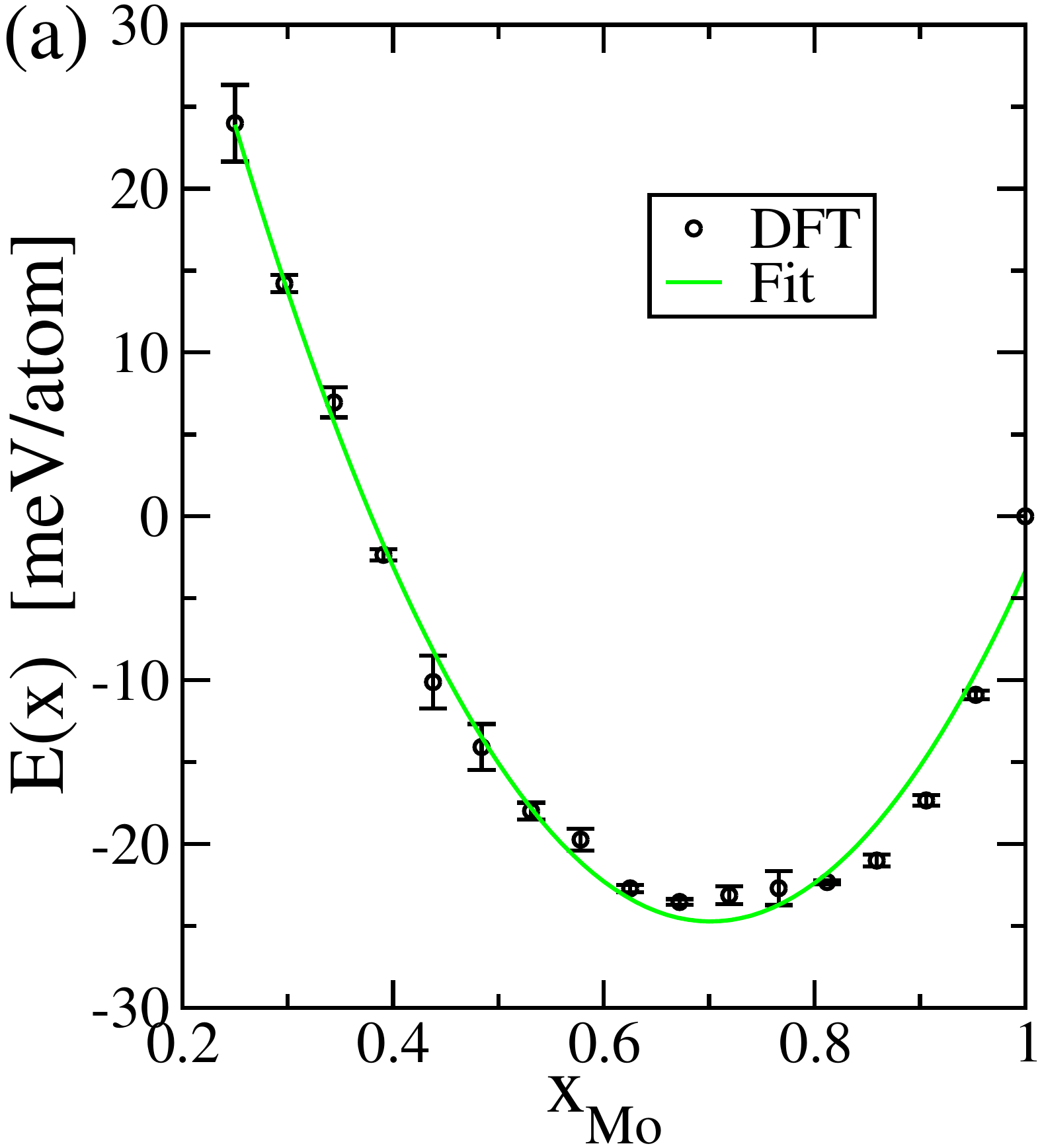}
\hspace{0.5cm}
\includegraphics[clip,width=1.8in]{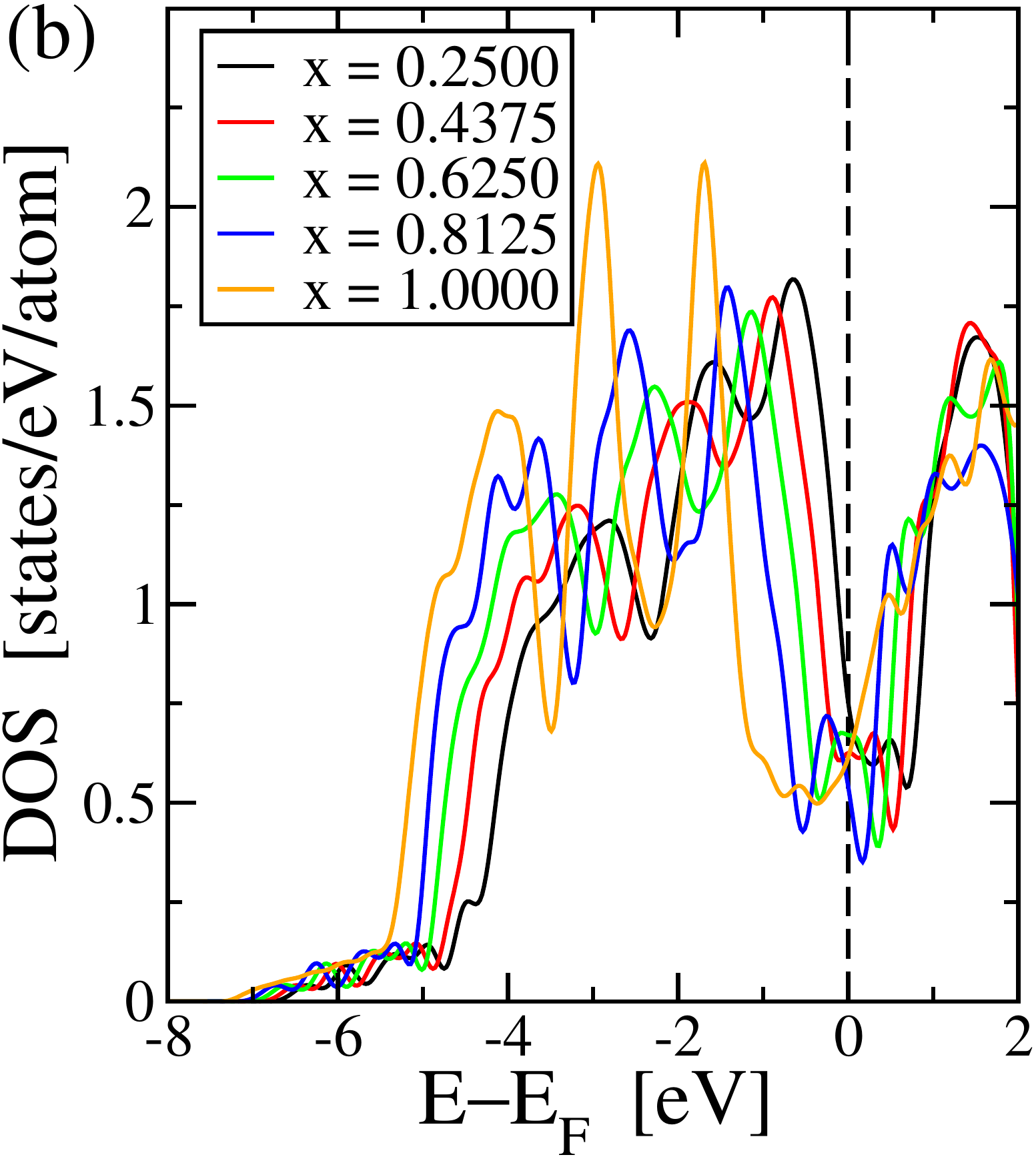}
\hspace{0.5cm}
\includegraphics[clip,width=1.8in]{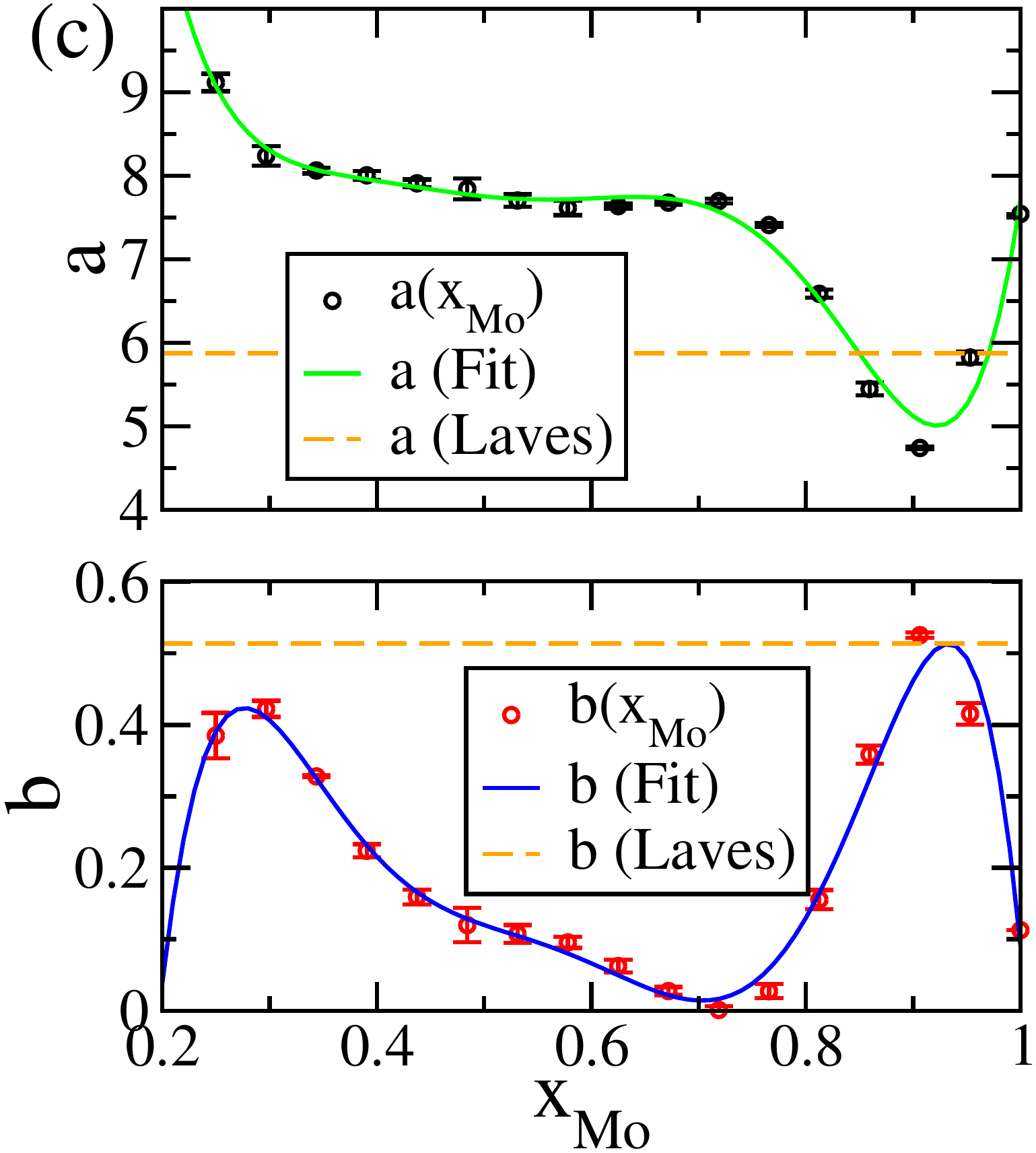}
\caption{\label{fig:x-series} (a) Formation enthalpies of BCC HEAs calculated from DFT at various compositions, $x_{\rm Mo}$. (b) Electronic densities of states. (c) Coefficients of electronic free energy.}
\end{figure*}

\begin{figure}
\includegraphics[clip,width=6in]{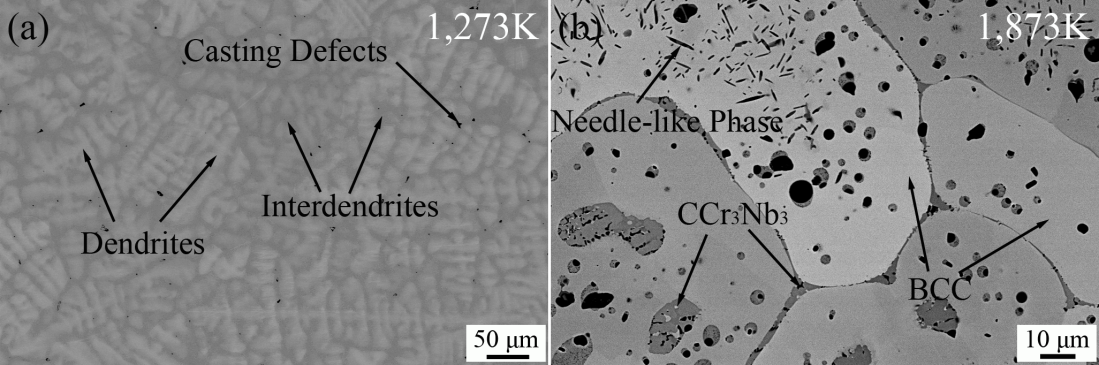}
\caption{\label{fig:sem} Electron micrographs of the (a) Cr-Mo-Nb-V sample following T = 1,273K anneal, and (b) following T = 1,873K.}
\end{figure}

\begin{figure}
\includegraphics[clip,width=6in]{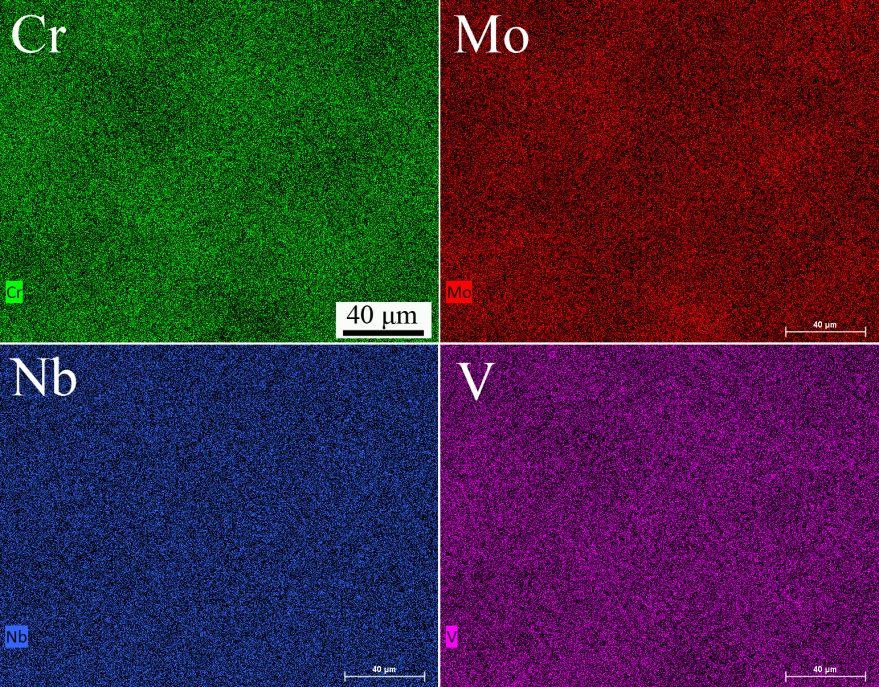}
\caption{\label{fig:edx-BCC} The chemical distribution of constitutive elements after annealing the T = 1,473K annealed sample at T = 1,673K for 12 hours.}
\end{figure}

\begin{figure}
\includegraphics[clip,width=6in]{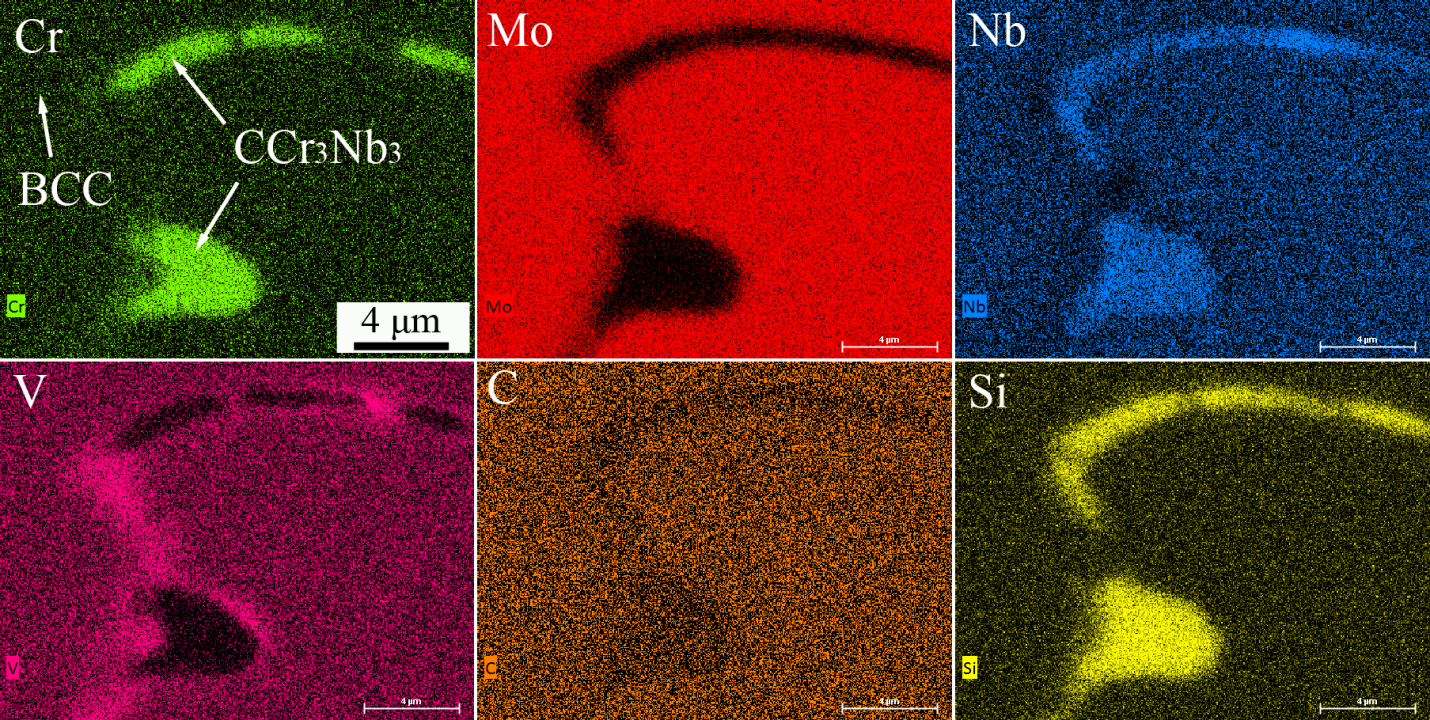}
\caption{\label{fig:edx-carbide} The chemical distribution of the BCC and CCr$_3$Nb$_3$ phases in the T = 1,873K-annealed state.}
\end{figure}

\begin{figure*}[h]
\includegraphics[clip,width=6in]{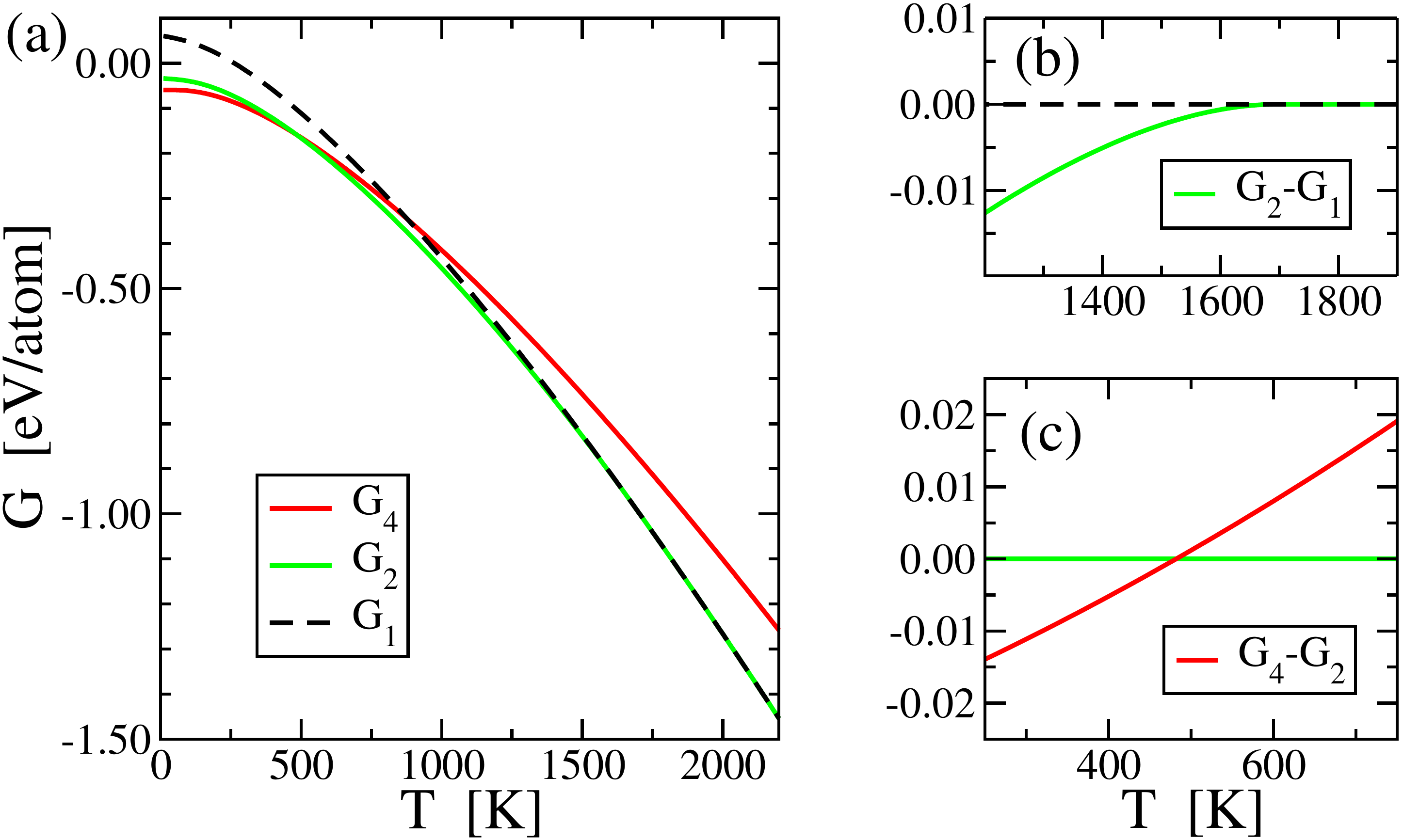}
\caption{\label{fig:G-all} Free energies for: $G_1$, a single-phase BCC high entropy alloy; $G_2$ a two-phase mixture of the BCC medium entropy alloy with the coexisting Laves phase; $G_4$ a four-phase mixture of the Laves phase with chemically ordered BCC-based structures.}
\end{figure*}

\begin{figure*}[h]
\includegraphics[clip,height=2.1in]{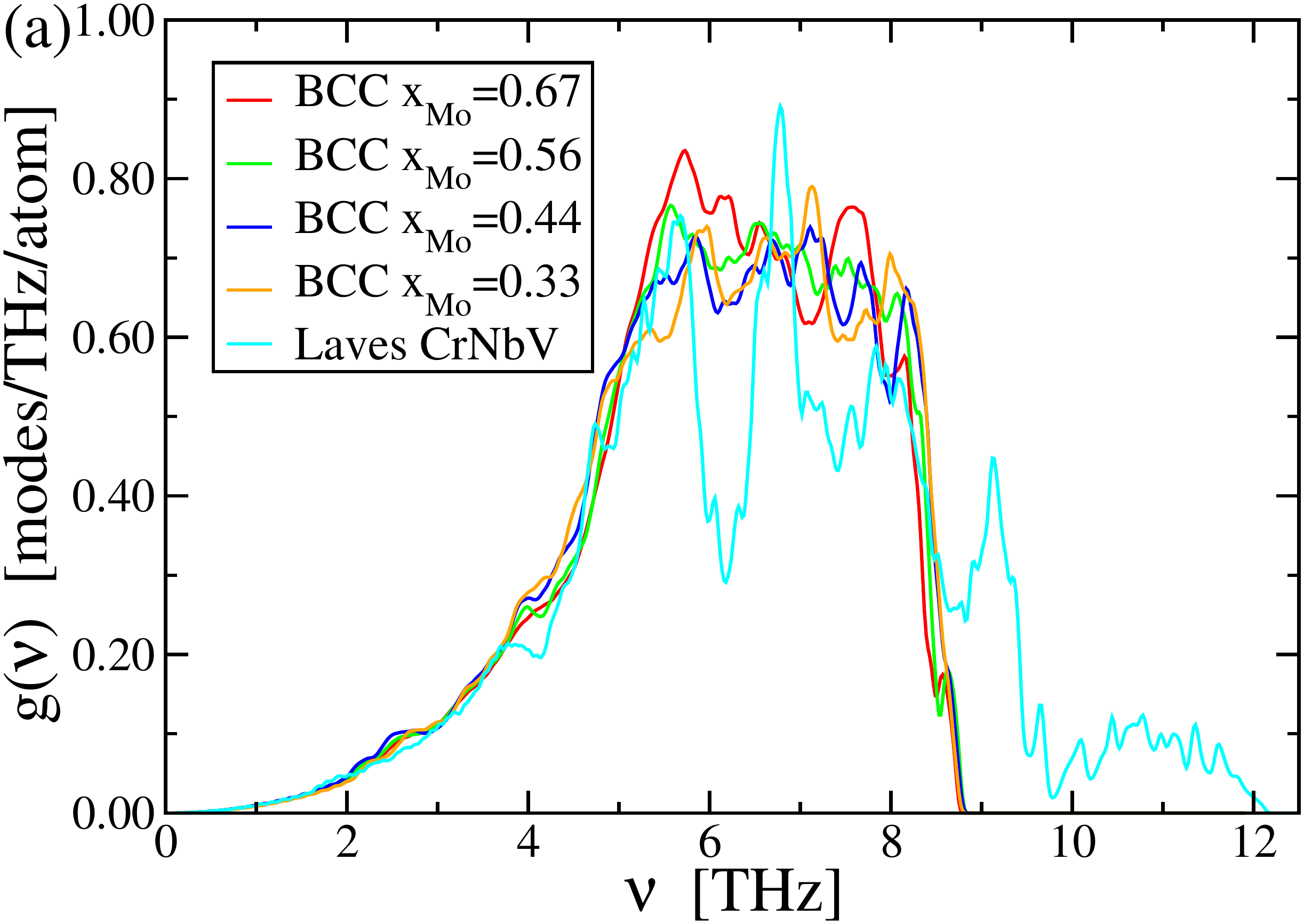}
\hspace{0.5cm}
\includegraphics[clip,height=2.1in]{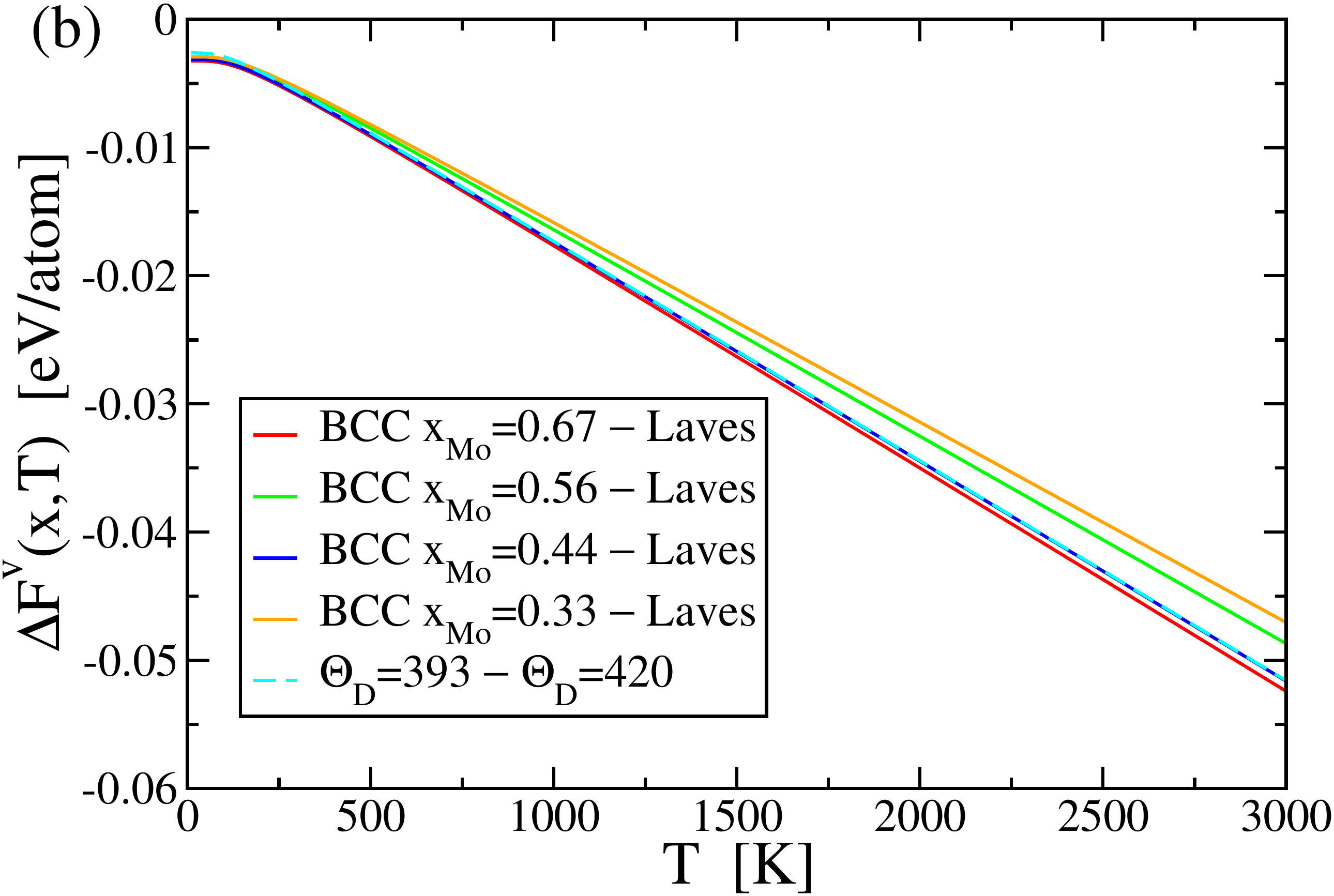}
\caption{\label{fig:vDOS} (a) Vibrational densities of states for BCC cells of varying composition compared with the CrNbV Laves phase. (b) Vibrational free-energy differences between BCC and Laves phases.}
\end{figure*}

\renewcommand{\baselinestretch}{1.0}

\begin{table}[tb]
\caption{\label{tab:Wyckoff} Structures of competing phases. Lattice constants are given in Angstroms, and Wyckoff positions in direct coordinates.}
\begin{tabular}{l|rr|lll}
\hline
Compound          & \multicolumn{5}{l}{Cr2V} \\
Pearson symbol    & \multicolumn{5}{l}{tI6} \\
Lattice constants & \multicolumn{5}{l}{a=2.87957, c=8.68376} \\
Space group       & \multicolumn{5}{l}{I4/mmm \#139} \\
\hline
Wyckoff  &   Cr  &   4e  &    0  &    0  &    0.17040 \\
Wyckoff  &    V  &   2b  &    0  &    0  &    1/2 \\
\hline
\hline
Compound & \multicolumn{5}{l}{MoNb}\\
Pearson & \multicolumn{5}{l}{symbol oC12}\\
Lattice constants & \multicolumn{5}{l}{a=13.80741, b=3.23206, c=4.55501}\\
Space group & \multicolumn{5}{l}{Cmmm \#65}\\
\hline
Wyckoff  &  Mo1  &   4h   &   0.17056   &   0   &   1/2\\
Wyckoff  &  Mo2  &   2a   &   0   &   0   &   0\\
Wyckoff  &  Nb1  &   2c   &   1/2   &   0   &   1/2\\
Wyckoff  &  Nb2  &   4g   &   0.66319   &   0   &   0\\
\hline
\hline
Compound & \multicolumn{5}{l}{Mo4V3}\\
Pearson symbol & \multicolumn{5}{l}{hR7}\\
Lattice constants & \multicolumn{5}{l}{a=4.34264, c=18.84620}\\
Space group & \multicolumn{5}{l}{R3m:H \#160}\\
\hline
Wyckoff &   Mo1  &   3a   &   0   &   0  &    0.00020\\
Wyckoff &   Mo2  &   3a   &   0   &   0  &    0.85778\\
Wyckoff &   Mo3  &   3a   &   0   &   0  &    0.57079\\
Wyckoff &   Mo4  &   3a   &   0   &   0  &    0.42837\\
Wyckoff &    V1  &   3a   &   0   &   0  &    0.28729\\
Wyckoff &    V2  &   3a   &   0   &   0  &    0.14128\\
Wyckoff &    V3  &   3a   &   0   &   0  &    0.71429\\
\hline
\hline
Compound & \multicolumn{5}{l}{CrNbV (disordered)}\\
Pearson symbol & \multicolumn{5}{l}{cF24}\\
Lattice constants & \multicolumn{5}{l}{a=7.039}\\
Space group & \multicolumn{5}{l}{Fd-3m \#227}\\
\hline
Wyckoff & Nb   &  8a &  0 & 0 & 0\\
Wyckoff & (Cr,V) & 16d & 5/8& 5/8& 5/8\\  
\hline
\hline
Compound & \multicolumn{5}{l}{CrNbV ordered}\\
Pearson symbol & \multicolumn{5}{l}{oI12}\\
Lattice constants & \multicolumn{5}{l}{a=4.94026, b=5.04117, c=7.02909}\\
Space group & \multicolumn{5}{l}{Imma \#74}\\
\hline
Wyckoff  &   Nb  &   4e    &    1/2 &     1/2  &    0.12378\\
Wyckoff  &   Cr  &   4a    &    0 &     0  &    0\\
Wyckoff  &    V  &   4d    &    1/4 &     1/2  &    3/4\\
\hline
\end{tabular}
\end{table}

\begin{table}[h]
\caption{\label{tab:carbide} Predicted structure of the carbide-impurity phase. Lattice constant is given in Angstroms, and Wyckoff positions are direct coordinates.}
\begin{tabular}{l|rr|lll}
\hline
Compound          & \multicolumn{5}{l}{CCr$_3$Nb$_3$} \\
Pearson symbol    & \multicolumn{5}{l}{cF112} \\
Lattice constants & \multicolumn{5}{l}{a=11.55} \\
Space group       & \multicolumn{5}{l}{$Fd\bar{3}m$ \#227} \\
\hline
Wyckoff  &    C  &  16d  &  1/4  &  1/2  &    1/4 \\
Wyckoff  &  Cr1  &  16c  &  1/4  &    0  &    1/4 \\
Wyckoff  &  Cr2  &  32e  & 0.453 & 0.797 &  0.453 \\
Wyckoff  &   Nb  &  48f  & 0.625 & 0.432 &  0.625 \\
\hline
\end{tabular}
\end{table}



\end{document}